\documentclass[proceedings]{JHEP3}

\PrHEP{PrHEP hep2001}
\conference{International Europhysics Conference on HEP}

\usepackage{epsfig,epic}

\newcommand{\Z}{\mathrm{Z}}

\newcommand{\Bss}       {\mathrm{B}^{(*)}_{\mathit{J}}}
\newcommand{\Bstar}     {\mathrm{B}^{*}     }
\newcommand{\B}         {\mathrm{B}         }

\newcommand{\Dsprimepm} {\mathrm{D}^{*\prime\pm}}
\newcommand{\Dsprimep}  {\mathrm{D}^{*\prime +}}
\newcommand{\Dsprime}   {\mathrm{D}^{*\prime}}

\newcommand{\Dstarpm}   {\mathrm{D}^{*\pm}  }
\newcommand{\Dstarp}    {\mathrm{D}^{*+}    }
\newcommand{\Dstar}     {\mathrm{D}^{*}     }
\newcommand{\Dzero}     {\mathrm{D}^{0}     }
\newcommand{\D}         {\mathrm{D}         }

\newcommand{\Km}        {\mathrm{K}^{-}     }
\newcommand{\pionp}     {\mathrm{\pi}^{+}   }
\newcommand{\pionm}     {\mathrm{\pi}^{-}   }

\title{Excited B and D Mesons at OPAL}

\author{\speaker{Kristian Harder}\\                       
        University of Hamburg/DESY, Notkestra\ss e 85,
        22607 Hamburg, Germany\\                           
        E-mail: \email{Kristian.Harder@desy.de}}           
      
      \abstract{Two recent OPAL publications dealing with spectroscopy
        of heavy-light mesons will be discussed here. In the charm
        sector, a search for a narrow radial excitation of the
        $\Dstarpm$ is performed. No signal is seen, and an upper
        limit of the production rate of narrow radial excitations
        close to the predicted mass of 2.629 GeV is derived.
        
        Orbitally excited $\Bss$ mesons are investigated in another
        analysis, where for the first time a measurement of their
        branching ratio into final states involving a $\Bstar$ is
        performed.  Attempts are made to separate the $\Bss$ signal
        into the four contributing resonances.  }

\begin{document}

\section{Introduction}

\FIGURE{\resizebox{0.49\textwidth}{!}{{\huge
\bf
 \begin{picture}(15.5,12)(0,1)

   \put(0,0){\framebox(15.5,13){ }}

   \linethickness{1mm}
   \put(0.5,7.5){\rotatebox{90}{mass (GeV/c$^2$)}}
   \put(  3, 0.5){\line(0,1){12}}
   \put(2.7, 1.5){\line(1,0){0.6}} \put(1.5, 1.3){5.2}
   \put(2.7, 3.5){\line(1,0){0.6}} \put(1.5, 3.3){5.4}
   \put(2.7, 5.5){\line(1,0){0.6}} \put(1.5, 5.3){5.6}
   \put(2.7, 7.5){\line(1,0){0.6}} \put(1.5, 7.3){5.8}
   \put(2.7, 9.5){\line(1,0){0.6}} \put(1.5, 9.3){6.0}
   \put(2.7,11.5){\line(1,0){0.6}} \put(1.5,11.3){6.2}

   \put( 4,2.35){\line(1,0){1.0}}  \put( 4.25,1.50){B}
   \put( 6,2.74){\line(1,0){1.0}}  \put( 6.20,1.89){B$^*$}
   \put( 8,6.88){\line(1,0){1.0}}  \put( 8.20,7.23){B$^*_0$}
   \put(10,7.07){\line(1,0){1.0}}  \put(10.20,7.32){B$_1$}
   \put(12,6.60){\line(1,0){1.0}}  \put(12.20,6.85){B$_1$}
   \put(14,6.69){\line(1,0){1.0}}  \put(14.20,6.99){B$^*_2$}
   \put( 4,8.33){\line(1,0){1.0}}  \put( 4.25,8.48){B$^\prime$}
   \put( 6,8.48){\line(1,0){1.0}}  \put( 6.20,8.63){B$^{*\prime}$}

   \put( 7.2,2.35){\line(1,0){7.8}}
   \put( 7.2,2.74){\line(1,0){7.8}}

   \thinlines
   \put( 4.5,8.33){\thicklines \dottedline{0.1}(0,0)(0,-5.97)}
   \put( 4.5,2.46){\vector(0,-1){0.1}} \put(3.5,5){$\pi\pi$}
   \put( 6.5,8.48){\thicklines \dottedline{0.1}(0,0)(0,-5.73)}
   \put( 6.5,2.85){\vector(0,-1){0.1}} \put(5.5,5){$\pi\pi$}
   \put( 8.5,6.88){\thicklines \dottedline{0.1}(0,0)(0,-4.49)}
   \put( 8.5,2.46){\vector(0,-1){0.1}} \put(8,5){$\pi$}
   \put(10.5,7.07){\thicklines \dottedline{0.1}(0,0)(0,-4.32)}
   \put(10.5,2.85){\vector(0,-1){0.1}} \put(10,5){$\pi$}
   \put(12.5,6.60){\vector(0,-1){3.85}} \put(12,5){$\pi$}
   \put(14.3,6.69){\vector(0,-1){4.33}} \put(13.7,5){$\pi$}
   \put(14.6,6.69){\vector(0,-1){3.94}} \put(14.7,5){$\pi$}
   \put(   6,2.74){\thicklines \dashline[+20]{0.2}(0,0)(-1,-0.39)}
   \put( 5.1,2.4){\vector(-2,-1){0.1}} \put(5,2.8){$\gamma$}
   \put(10,1.5){\thicklines \dottedline{0.1}(0,0)(0.9,0)}
   \put(10.9,1.5){\vector(1,0){0.1} \Large strong (S-wave)}
   \put(10,0.9){\vector(1,0){1.0} \Large strong (D-wave)}
   \put(10,0.29){\thicklines \dashline[+40]{0.2}(0,0)(0.9,0)}
   \put(10.9,0.3){\vector(1,0){0.1} \Large electromagnetic}
   \thicklines

   \put( 4,10.50){$\overbrace{\hspace*{3cm}}$}
   \put( 8,10.50){$\overbrace{\hspace*{7cm}}$}
   \put( 4.8,11){L=0}
   \put(10.1,11){L=1 (B$^{(*)}_J$)}

   \put( 8,8.50){$\overbrace{\hspace*{3cm}}$}
   \put(12,8.50){$\overbrace{\hspace*{3cm}}$}
   \put( 8.6,9){\Large $j_q=1/2$}
   \put(12.6,9){\Large $j_q=3/2$}
 \end{picture}
}
}
        \caption{\label{fig-spectra} Spectrum of $\B$ mesons.}  }

The spectra of mesons consisting of one heavy and one light quark can
be described perturbatively in the framework of Heavy Quark Effective
Theory (HQET). Precise predictions of the masses of excited mesons
have been made. Their experimental verification will help tune HQET
phenomenology and thus improve our general understanding of QCD. The
spectrum of $\B$ mesons as expected from one such prediction
\cite{ref-spectra} is displayed in Figure~\ref{fig-spectra}.  The
excited states are expected to decay by strong interaction, mainly via
emission of one or two pions. Allowed decay modes are
depicted by arrows. Two-pion decays of each orbital excitation to both
members of the ground state doublet are also expected, but probably
phase-space suppressed. Where kinematically possible, also decays
involving a $\rho$ meson might contribute. Decays of radial
excitations that proceed in two steps with an intermediate orbital
excitation are also allowed. A very similar spectrum is predicted for
$\D$ mesons, but the mass splittings within the doublets are larger
there due to the smaller charm quark mass.

Experimental access to excited heavy-light mesons turned out to be
severely constrained at present and past experiments due to limited
mass resolution, large background, and insufficiently large candidate
samples. The existing results on excited heavy-light mesons are
therefore partially inconclusive, and even contradictory in several
cases (see concluding remarks in Sections \ref{sec-dsp} and
\ref{sec-bss}).

Recent work on heavy-light meson spectroscopy with LEP1 data
contributed by the OPAL collaboration will be discussed in this paper.
A search for a narrow $\Dsprime$, the first radial excitation of the
$\Dstar$ meson, is presented in Section~\ref{sec-dsp}
\cite{ref-opaldsp}.  Section~\ref{sec-bss} contains a summary of the
results obtained by OPAL in the investigation of $\Bss$ mesons
\cite{ref-opalbss}, where $\Bss$ is a common notation for the four
orbitally excited $\B$ mesons with orbital angular momentum $\rm L=1$.
A brief comparison of the OPAL results with contributions by other
experiments will be included in both sections.

\section{A Search for a Narrow Radial Excitation of
         the \boldmath $\Dstar$ \unboldmath Meson}
\label{sec-dsp}

The DELPHI collaboration reported an observation of a narrow
($<15\thinspace\rm MeV/c^2$) resonance in $\Dstarp\pionp\pionm$ final
states\footnote{Charge conjugates are always implied.} a couple of
years ago \cite{ref-delphidsp}, whose mass coincided very well with
the predicted mass of the first radial excitation of the $\Dstar$
meson.  Despite this agreement, the observation was a surprise because
a much larger $\Dsprime$ width was favoured: The
$\Dsprimep\to\Dstarp\pionp\pionm$ decay is most likely dominated by
the S-wave contribution, which usually leads to widths of the order of
$100\thinspace\rm MeV/c^2$ in comparable systems. The association of
the observed resonance with $\Dsprime$ has therefore been questioned,
despite the lack of good alternative explanations
\cite{ref-theoryresp}.

The interesting result seen by DELPHI has triggered a similar analysis
at OPAL \cite{ref-opaldsp}, where $\Dstarp\pionp\pionm$ combinations
are looked at in search of any narrow resonant structure with a mass
close to both DELPHI observation and HQET $\Dsprime$ prediction. The
$\Dstarp$ mesons are reconstructed in their decay chain
$\Dstarp\to\Dzero\pionp$, $\Dzero\to\Km\pionp$. A combination of
$\Dstarp$ candidates with two pions results in the desired $\Dsprimep$
candidates.

\begin{figure}
{\center
\epsfig{file=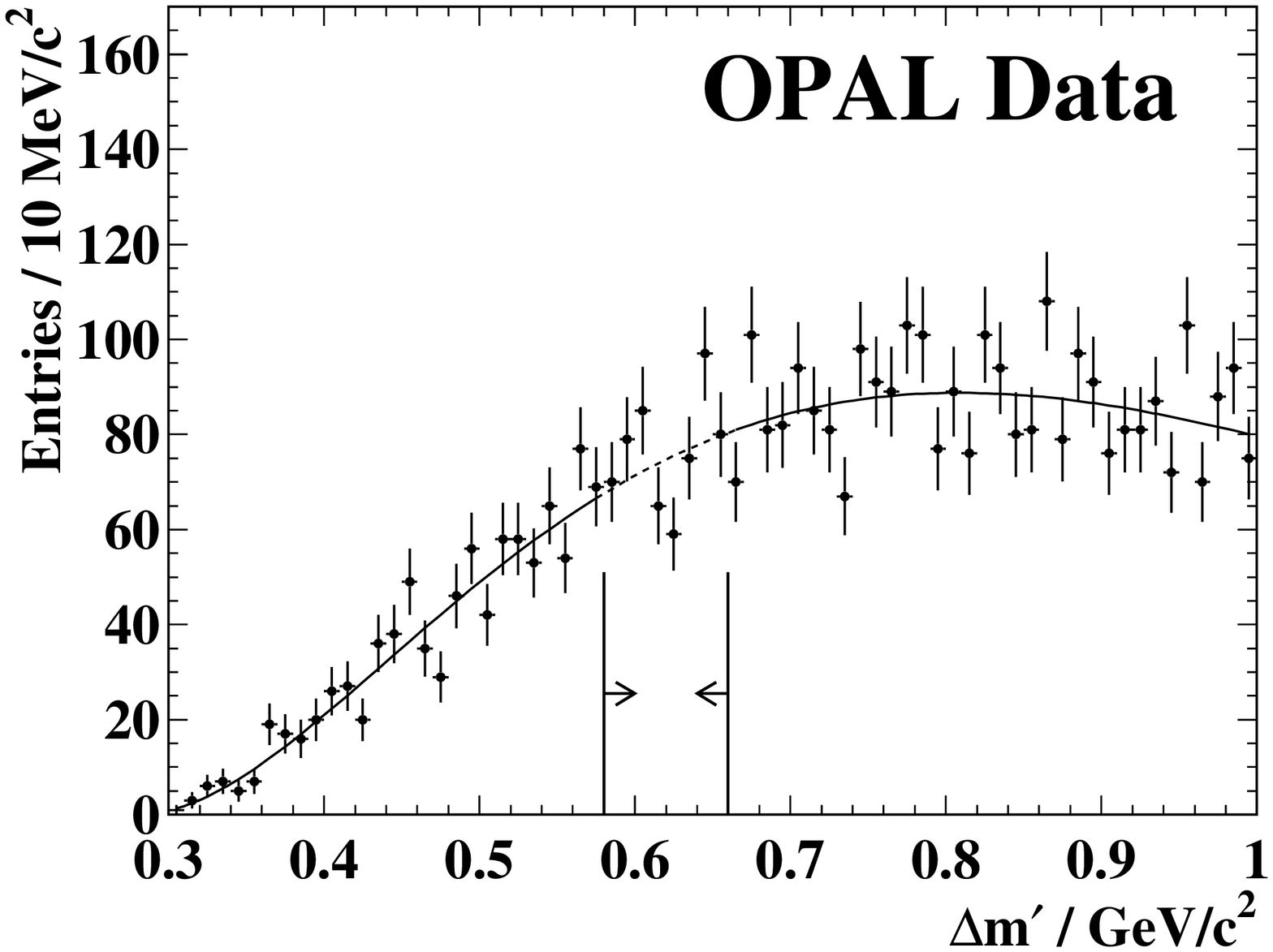,width=0.45\textwidth}
\hspace{0.5cm}
\epsfig{file=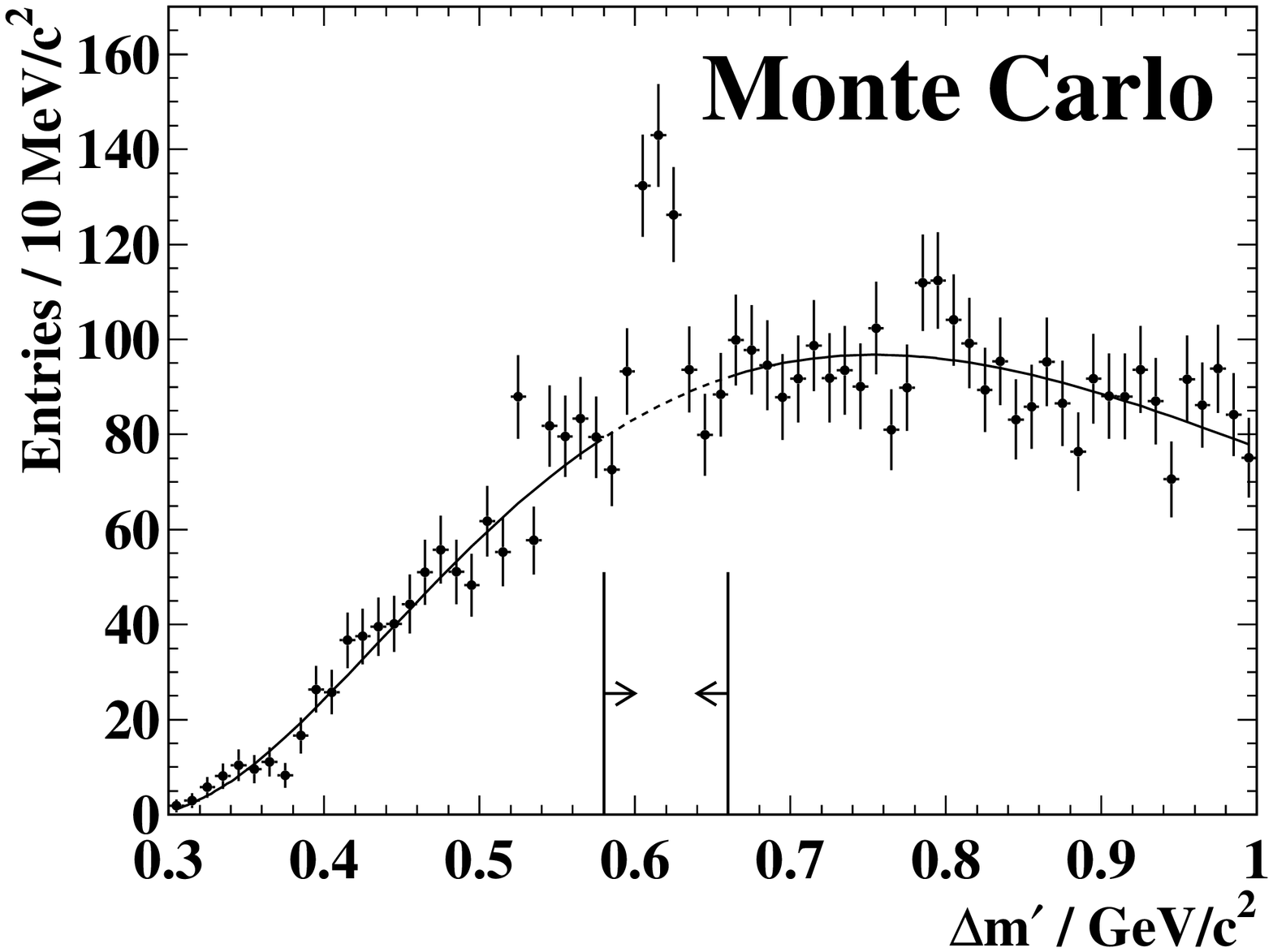,width=0.45\textwidth}
\caption{\label{fig-opaldsp} Mass spectrum
  of $\Dstarp\pionp\pionm$ candidates seen by OPAL (left). No narrow
  resonance is found. An upper limit on the production rate of a
  possible narrow radial excitation is calculated for a mass inside
  the mass window represented by the arrows. The right plot shows the
  corresponding spectrum for a Monte Carlo simulation including a
  resonance similar to the one reported by DELPHI.}}
\end{figure}

Figure~\ref{fig-opaldsp} shows the mass spectrum of
$\Dstarp\pionp\pionm$ combinations found by OPAL. No sign of a narrow
resonance is seen in a wide mass region around the signal reported by
DELPHI. A clear peak over the non-resonant background was expected
from Monte Carlo including a resonance with parameters adjusted to the
DELPHI observation.

In absence of any evidence for a signal in the investigated mass
range, an upper limit on the production rate of narrow radial
excitations close to the predicted mass of 2.629 GeV is derived:
\[
f(\Z \to \Dsprimepm(2629)) \times {\rm Br}(\Dsprimep \to \Dstarp \pionp
\pionm) < 3.1 \times 10^{-3} \ \ \mathrm{(95\%\ C.L.)}
\]
This result does not depend significantly on specific properties of
radial excitations and is thus valid also for other possible narrow
resonances in the mass region in question. No production of narrow
$\Dsprime$ or similar resonances in primary $\rm c\bar{c}$ events or
$\rm b\bar{b}$ events is observed, and separate limits for these two
cases are obtained:
\[
    f({\rm c} \to  \Dsprimep(2629)) \times {\rm Br}(\Dsprimep \to \Dstarp \pionp
    \pionm) < 0.9 \times 10^{-2} \ \ (95\%\ \mathrm{C.L.})
\]
\[
    f({\rm b} \to  \Dsprimep(2629)) \times {\rm Br}(\Dsprimep \to \Dstarp \pionp
    \pionm) < 2.4 \times 10^{-2} \ \ (95\%\ \mathrm{C.L.})
\]
This non-observation of the resonance seen by DELPHI is supported by
other, as yet preliminary analyses by CLEO \cite{ref-cleodsp} and ZEUS
\cite{ref-zeusdsp}.

\section{Investigation of the Decay of Orbitally Excited B Mesons}
\label{sec-bss}

The orbital excitations $\Bss$ of the $\B$ meson spectrum are well
established; however, the large mass of the $\B$ quark leads to only
small mass splittings between the individual states. These mass
splittings are smaller than or of about the same size as the widths of
the individual $\Bss$ states. Therefore the resonances overlap, and it
is experimentally very difficult to distinguish them.

The main handle to improve knowledge about the $\Bss$ substructure is
a close investigation of their decay. Figure~\ref{fig-spectra} shows
that mainly the decays $\Bss\to\Bstar\pi$ and $\Bss\to\B\pi$ are
expected. In particular, three out of the four states can only decay
into exactly one of those two final states. A separation of
$\Bss\to\Bstar\pi$ from $\Bss\to\B\pi$ decays would thus be a very
useful tool to gain insight into the the composition of the $\Bss$
spectrum. To distinguish these decays is very difficult, because the
$\Bstar$ mesons decay to $\B\gamma$, and thus the only visible
difference in the final state is a low-energetic photon. Many other
sources for low energy photons lead to a background level large enough
to render any attempt to reconstruct a full $\B\gamma\pi$ final state
virtually impossible for current experiments.

OPAL \cite{ref-opalbss} reconstructs $\Bss$ candidates by first
tagging $\rm b\bar{b}$ events looking at $\B$ decay vertices, high
$p_t$ leptons, and jet shapes. All tracks with suitable kinematic
properties are then combined to form inclusive $\B$ candidates. With
an additional pion, one obtains a mass distribution with a clear
excess caused by $\Bss\to\B\pi(X)$ decays (see Fig.~\ref{fig-bss}a).
Here, $X$ represents possible additional final state particles like
another pion from two-pion transitions which are expected to be
suppressed, but not excluded. Also, $X$ includes photons from
$\Bss\to\Bstar\pi\to\B\gamma\pi$ cascade decays.

A good description of the background is mandatory to obtain useful
results after background subtraction. All important background sources
are studied independently at OPAL by creating samples that are
enriched in the respective type of background. The shape of these
samples is then compared to corresponding data samples, and the
relative size of the background samples in Monte Carlo is weighted to
match the data best. The resulting background-subtracted distribution
is shown in Figure~\ref{fig-bss}b. Because the efficiency to
reconstruct $\Bss$ decays is mass-dependent, an efficiency-correction
is applied to obtain the $\Bss$ mass distribution
(Fig.~\ref{fig-bss}c).

\EPSFIGURE[t]{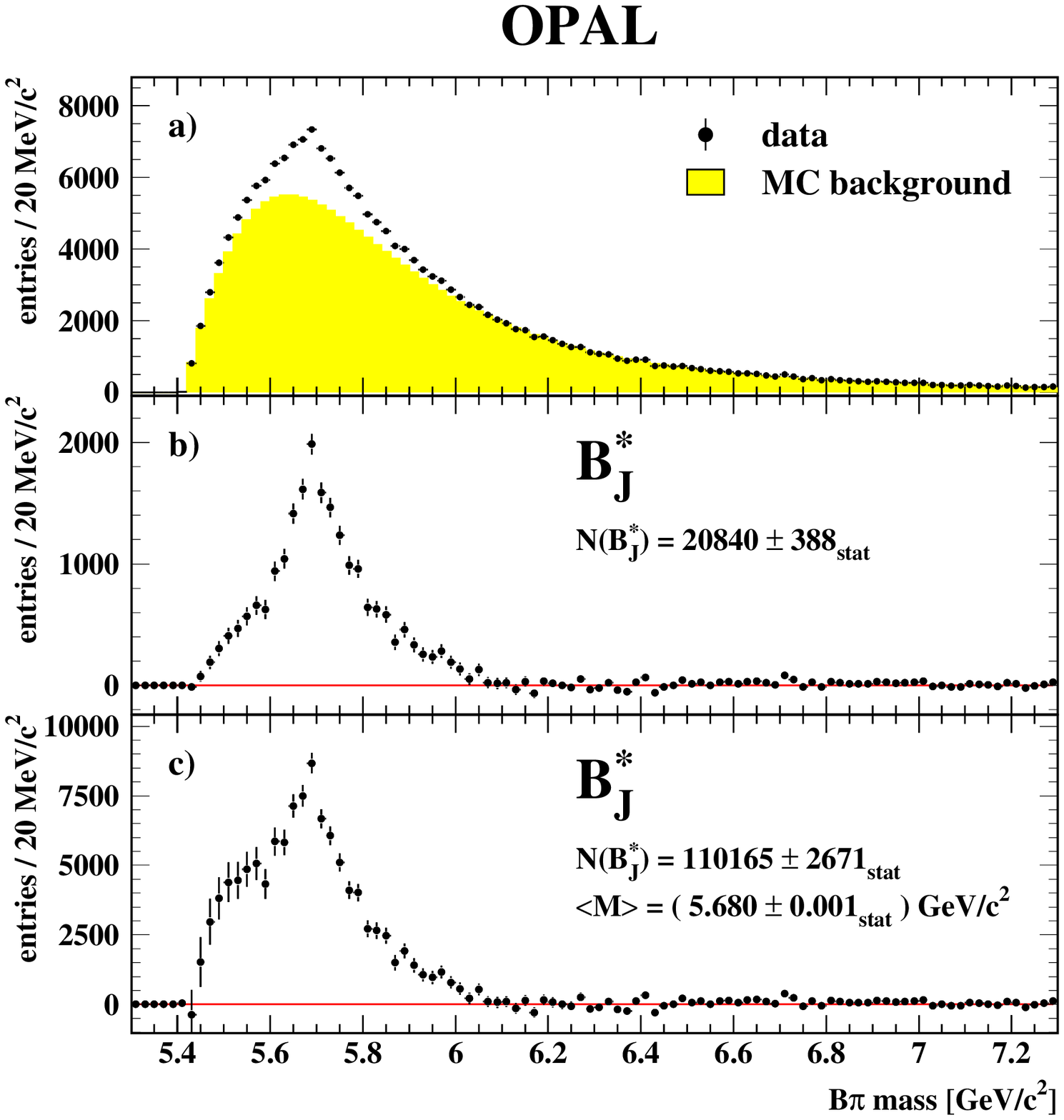,width=0.59\textwidth}{\label{fig-bss} Invariant
  mass distribution of $\B\pi$ combinations in OPAL data (top). A
  clear signal from $\Bss\to\B\pi(X)$ decays is visible. Below are the
  background-subtracted (centre) and efficiency-corrected mass
  distributions (bottom).}

A combination of $\B\pi$ candidates with photons to obtain
$\Bss\to\Bstar\pi\to\B\gamma\pi$ candidates would completely dilute
the $\Bss$ signal due to bad mass resolution and large background. A
different approach is therefore used by OPAL. The number and quality
of photon candidates that can be combined with a $\B\pi$ candidate to
form a $\Bstar\pi$ candidate with acceptable properties is evaluated
into a $\Bstar$ weight. $\B\pi$ combinations from $\Bss\to\Bstar\pi$
decays tend to have a higher $\Bstar$ weight than $\B\pi$ combinations
from $\Bss\to\B\pi$ decays.  By cutting on this weight, the sample of
$\B\pi$ candidates can thus be divided into two subsamples: One is
enriched in $\B\pi$ combinations from $\Bss\to\Bstar\pi$ decays, the
other is enriched in $\Bss\to\B\pi$ decays. However, no specific
photon candidate is assigned to the $\B\pi$ candidates in the former
sample. The invariant mass of a $\Bss\to\Bstar\pi$ candidate is
calculated as the invariant mass of the $\B\pi$ distribution, plus the
world average mass difference of the $\Bstar$ and $\B$ mesons.

The efficiencies to reconstruct $\Bss\to\B\pi$ and $\Bss\to\Bstar\pi$
decays are different for the two ($\Bstar$-enriched and
$\Bstar$-depleted) samples. This allows to calculate for the first
time the branching ratio of $\Bss$ decays to final states involving a
$\Bstar$:
\[
 {\rm Br}(\Bss\to\Bstar\pi(X))=0.85^{+0.26}_{-0.27}(stat.)\pm0.12(syst.)
\]
Due to the inclusive character of the $\B$ meson reconstruction, a
final state with more light particles ($X$) than the reconstructed pion
cannot be distinguished from a pure $\Bstar\pi$ final state. Specifically,
$\Bss\to\Bstar\pi\pi$ decays might contribute.

The goal to disentangle the contributions of the four individual
resonances to the $\Bss$ peak remains ambitious. A simultaneous fit to
the background-subtracted and efficiency-corrected mass distributions
of both $\B\pi$ samples is performed in the HQET framework to obtain
measurements of some parameters at the price of fixing others at their
predicted values.  Still, parts of the fit are inherently unstable,
and the systematic errors are large. Furthermore, similar fits by
ALEPH \cite{ref-alephbss} and L3 \cite{ref-l3bss} lead to different
conclusions on the masses of the broad $\Bss$ resonances, the
existence of radial excitations, and the contribution of di-pion
transitions. The most reliable OPAL fit results are the mass of the
narrow $\rm B_1$, found to be $M(\rm
B_1(3/2))=(5.738^{+0.005}_{-0.006}\pm0.007) {\rm GeV}/c^2$, and its
width of $\rm \Gamma(B_1(3/2))=(18^{+15+29}_{-13-23}) {\rm MeV}/c^2$.

\section{Conclusion}

Years after the end of the LEP1 program, heavy flavour spectroscopy
with LEP1 data is still an active field.  OPAL has recently
contributed new results on radially excited $\D$ mesons and on
orbitally excited $\B$ mesons. However, in both cases comparison with
the results obtained by other collaborations shows that the results
obtained at LEP are not sufficiently clear to solve all questions we
would like to answer in this field: In the charm sector the DELPHI and
OPAL collaborations disagree on whether a narrow radial excitation of
the $\Dstar$ mesons is present or not. $\Bss$ mesons are well
established, but attempts to separate the $\Bss$ peak into the
contributions of the four individual $\Bss$ states (and possibly
radial $\B$ excitations) lead to contradictory results among the
ALEPH, L3 and OPAL collaborations. Using a novel approach to separate
$\Bss\to\Bstar\pi(X)$ from $\Bss\to\B\pi(X)$ decays, OPAL performs the first
measurement of the branching ratio ${\rm Br}(\Bss\to\Bstar\pi(X))$.

\end{document}